\documentclass[journal]{vgtc}                

\usepackage{url}

\PassOptionsToPackage{bookmarks=false}{hyperref}

\ifpdf
  \pdfoutput=1\relax                   
  \usepackage{graphicx}                
  \DeclareGraphicsExtensions{.pdf,.png,.jpg,.jpeg} 
\else
  \usepackage{graphicx}                
  \DeclareGraphicsExtensions{.eps}     
\fi%

\graphicspath{{figures/}{pictures/}{images/}{./}} 
\usepackage{balance}
\usepackage{microtype}                 
\PassOptionsToPackage{warn}{textcomp}  
\usepackage{textcomp}                  
\usepackage{mathptmx}                  
\usepackage{times}                     
\usepackage{cite}                      
\usepackage{tabu}                      
\usepackage{booktabs}                  

\onlineid{0}

\usepackage[table]{xcolor}
\usepackage{color, colortbl}
\usepackage{fancyhdr}
\usepackage{ctable}
\definecolor{Gray}{gray}{0.9}
\usepackage{verbatim}
\usepackage{ragged2e}
\usepackage{amsmath}

\usepackage{array}
\newcolumntype{L}[1]{>{\raggedright\let\newline\\\arraybackslash\hspace{0pt}}m{#1}}
\newcolumntype{C}[1]{>{\centering\let\newline\\\arraybackslash\hspace{0pt}}m{#1}}
\newcolumntype{R}[1]{>{\raggedleft\let\newline\\\arraybackslash\hspace{0pt}}m{#1}}

\usepackage{enumitem}

\vgtcpapertype{application/design study}

\title{Humane Visual AI: Telling the Stories Behind a Medical Condition}

\author{Wonyoung So, Edyta P. Bogucka, Sanja \v{S}\'{c}epanovi\'{c}, Sagar Joglekar, Ke Zhou, and Daniele Quercia}
\authorfooter{
\item
 Wonyoung So is with Massachusetts Institute of Technology, Cambridge, MA (US). E-mail: wso@mit.edu.
\item
 Edyta P. Bogucka is with  Chair of Cartography, Technical University of Munich (DE). E-mail: e.p.bogucka@tum.de.
   \item
 Sanja \v{S}\'{c}epanovi\'{c} is Nokia Bell Labs, Cambridge (UK), E-mail: sanja.scepanovic@nokia-bell-labs.com.
  \item
 Sagar Joglekar is Nokia Bell Labs, Cambridge (UK), E-mail: sagar.joglekar@nokia-bell-labs.com.
 \item
 Ke Zhou is Nokia Bell Labs, Cambridge (UK), E-mail: ke.zhou@nokia-bell-labs.com.
\item
 Daniele Quercia is Nokia Bell Labs, Cambridge (UK), CUSP King's College London. E-mail: daniele.quercia@nokia-bell-labs.com.
}

\shortauthortitle{Wonyoung So \MakeLowercase{\textit{et al.}}: Humane Visual AI: Telling the Stories Behind a Medical Condition}

\teaser{
  \centering
  \includegraphics[width=0.88\linewidth]{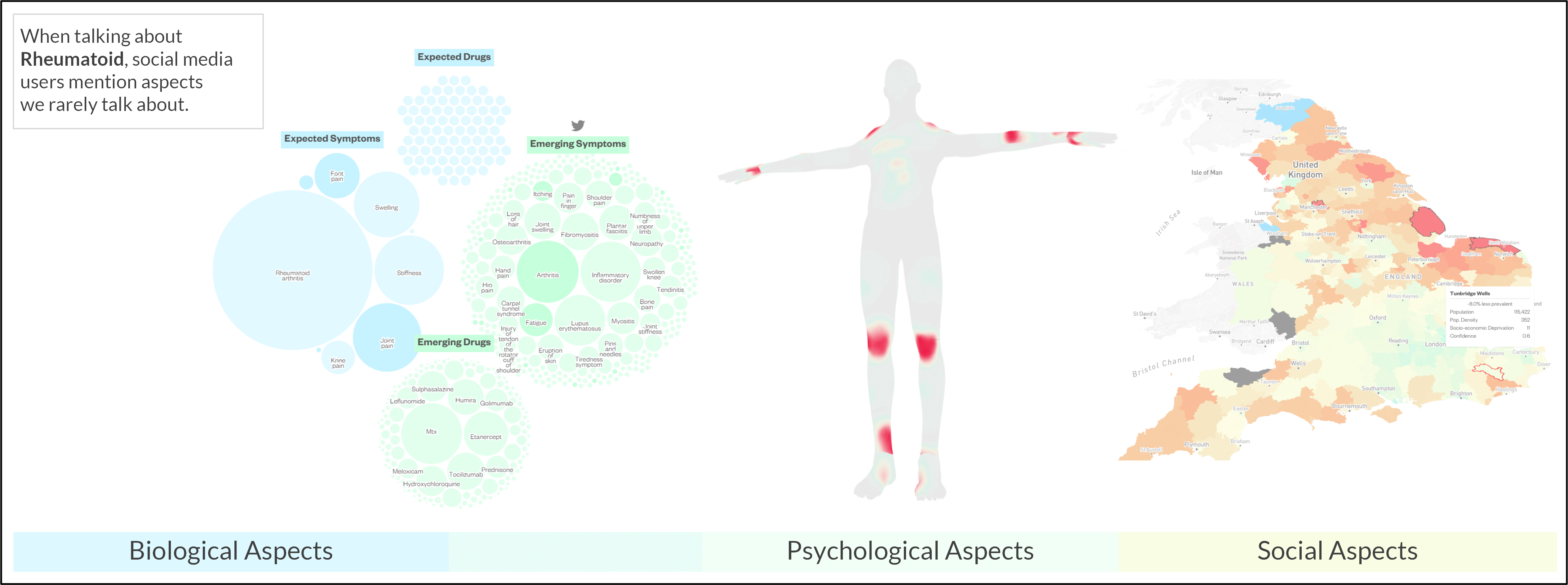}
  \caption{Narrating the story behind a medical condition through biological, psychological, and socio-environmental lens (i.e., through the bio-psycho-social model).}
  \label{fig:teaser}
}

\abstract{
A biological understanding is key for managing medical conditions, yet psychological and social aspects matter too.  The main problem is that these two aspects are hard to \emph{quantify} and inherently difficult to \emph{communicate}. To \emph{quantify}  psychological aspects, this work mined around half a million Reddit posts in the sub-communities specialised in $14$ medical conditions, and it did so with a new deep-learning framework. In so doing, it was able to associate mentions of medical conditions with those of emotions.  To then \emph{quantify}   social aspects, this work designed a probabilistic approach that mines open prescription data from the National Health Service in England to compute the prevalence of drug prescriptions, and to relate such a prevalence to census data. 
To finally {\emph{visually}} \emph{communicate} each medical condition’s biological, psychological, and social aspects through storytelling, we designed a narrative-style layered Martini Glass visualization.
In a user study involving $52$ participants, after interacting with our visualization, a considerable number of them changed their mind on previously held opinions: 10\% gave more importance to the psychological aspects of medical conditions, and 27\%  were more favourable to the use of social media data in healthcare, suggesting the importance of persuasive elements in interactive visualizations.
} 


\CCScatlist{
  \CCScatTwelve{Human-centered computing}{Human Computer Interaction (HCI)}{Interaction paradigms}{Web-based interaction}
  \CCScatTwelve{Human-centered computing}{Visualization}{Empirical studies in visualization}{}
}

\begin{document}


\firstsection{Introduction}

\maketitle
The communication between patient and healthcare professional often focuses on the biological aspects of a medical condition (e.g., high temperature or pain) and overlooks its psychological and social aspects (e.g., patient feeling sad or lacking control due to recent difficult events in their social life). Yet these two latter aspects are considered to be equally important by the modern ``bio-psycho-social model'' in healthcare ~\cite{engel_need_1977,fava2007biopsychosocial,suls2004evolution}.  
{
Social and psychological aspects such as fear, rage, neglect, attachment, mental well-being, life course, social capital, healthy lifestyle, neighborhood disadvantage, and perceived discrimination are all shown to have important effects on chronic conditions such as obesity, coronary heart disease, and skin disease \cite{engel_need_1977,ricci2018obesity},
as well as on functional disorders \cite{van2016biopsychosocial,pick2019emotional}, 
and medically unexplained symptoms \cite{engel_need_1977,fava2007biopsychosocial}.
} 
 {After reviewing the literature, 
we found the most important psychological and social aspects to be the following: i) the patients' \emph{emotions} (such as fear or trust)  \cite{engel_need_1977,consedine2007role},
ii) their \emph{psychological symptoms} (such as depression and anxiety) \cite{engel_need_1977,bayat2011symptoms}, 
and iii) the \emph{socio-economic status} of the areas where they live (such as neighbourhood disadvantage and social network) \cite{engel_need_1977,cutler2008socioeconomic,aiello2019large}. 
}

The main problem is that psychological and social aspects are: 1) hard to quantify; and 2) inherently difficult to communicate. First, they are hard to quantify because they have not been systematically recorded. To see why, consider the psychological aspects. There is no protocol in National Health Services that mandates the systematic recording of patients' feelings and emotions. Yet, in recent years, individuals have increasingly discussed their medical conditions on sites such as: \emph{PatientsLikeMe} where they connect with others suffering from the same condition; \emph{AskaPatient} where they share experiences with drugs; and the general-purpose \emph{Reddit} and \emph{Twitter}. It turns out that people tend to reveal even more personal experiences online than they would usually do in face-to-face interactions with their doctors (e.g., they discuss emotions, fears, and life events) -- this is what has been called the disinhibition effect \cite{suler2004online}. The problem with social media text is that it is unstructured and, as such, it is hard to process for the reliable extraction of medical conditions. The quantification of social aspects is challenging too -- it is hard to relate socio-economic data (e.g., census data) to health outcomes. 

Second, even if one were to be able to quantify psychological and social aspects, these would still be inherently difficult to communicate. That is simply because these aspects are part of a multi-faceted bio-psycho-social model. Multi-dimensional data is hard to visualize in engaging ways~\cite{big-data-engagement}, and this represents an active research area~\cite{big-data-viz-advances, big-data-health, geo-big-data}. 
 {
Visual storytelling was suggested to be the next-level approach in healthcare communication \cite{mccurdy2016visual,briant2016power}.} 
 

This work tackled the two challenges of both \emph{quantifying} and \emph{visually communicating} a medical condition's psychological and social aspects. In so doing, it made three main contributions:

\begin{itemize}[leftmargin=*]
\itemsep0em
\item[C1:] By considering around half a million Reddit posts in the sub-communities specialised in $14$ medical conditions, for each condition, we \emph{quantified} its psychological and social aspects by using (Section~\ref{sec:quantify}):  a) our new deep-learning framework to extract medical entities from the posts with state-of-the-art accuracy \cite{scepanovic_martin-lopez_quercia_2020}; 
b) a widely adopted dictionary-based approach to extract emotions from the posts; and c) a probabilistic approach that mines open prescription data from the National Health Service in England to compute the prevalence of drug prescriptions related to the condition. 

\item[C2:] We \emph{visually communicated} each medical condition's biological, psychological, and social aspects through a layered Martini Glass visualization (Section \ref{sec:communicate}).


\item[C3:] We evaluated our visualization and found that it had persuasive effects (Section \ref{sec-evaluation}). In a user study, after interacting with our visualization, a considerable number of our $52$ participants changed their mind:
10\% gave more importance to the psychological and social aspects of medical conditions; and 27\%  were more favourable to the use of social media data, and 10\% to the use of AI in healthcare, suggesting the importance of persuasive elements in interactive visualizations.

\end{itemize}

The resulting visualization is available on the project's page \break  \url{http://humane-ai.social-dynamics.net}.

\section{Related Work}\label{sec:relatedwork}
There have been two lines of past research related to ours. The first concerns general ways of quantifying health outcomes. The second concerns visualizing multi-faceted health aspects.

\vspace{3pt}\noindent\textbf{General ways of quantifying health outcomes.}
One of the primary tasks in mining health discussions is to {extract medical mentions}. Initial approaches to this task were \textit{keyword and lexicons-based} \cite{leaman2010towards,gonzalez2017capturing,park2017tracking}. They worked reasonably well on formal medical text but had well-known limitations when applied to social media text: they failed to capture the variability of informal language, diversity of users' symptom expressions, and spelling mistakes \cite{cohen2005survey,paul2016social}. 
The approaches that dealt better with such limitations were based on \textit{machine learning}, such as Conditional Random Fields (CRFs)~\cite{nikfarjam2015pharmacovigilance,lee2015enhanced} and on \textit{deep learning}, such as Recurrent Neural Networks (RNNs) \cite{tutubalina2017combination,scepanovic2020medred}.
\textit{Dictionary-based approaches} were used to extract linguistic features from posts, which were then shown to relate to chronic disease health outcomes, both on Twitter \cite{Culotta204twitter} and Reddit \cite{de2014mental}. Psychological and social aspects (such as positive emotions, work-related issues, or stigma) were studied on Reddit \cite{park2018examining} and on Twitter \cite{robinson2019measuring}. The drug prescription data from the general practitioners across England were used as a \textit{proxy for prevalence of diseases}: examples of such uses include the health domains of diabetes ~\cite{aiello2019large}, influenza~\cite{perrotta2019spatio}, and opioid addiction \cite{curtis2019opioid}.

\vspace{3pt}

\noindent {\textbf{Visualizing multi-faceted information: Dashboard and Narrative Visualization.} 
}
 {Dashboards have been one of the most popular choices for visualizing medical data, as they can consolidate multi-faceted clinical conditions in one screen \cite{elias_exploration_2011, lanchantin_deep_2016, faiola_supporting_2015}. However, clinical dashboards mainly supported medical professionals \cite{khairat2018impact}, and, more generally,  dashboard visualizations have been found to have two main weaknesses: information overload given the single screen \cite{wilbanks2014review}, and a significant learning curve \cite{khairat2018impact}. }


 {An alternative visualization technique for delivering complex information is the \emph{narrative visualization}. The idea is to create a story based on the results of the data analysis. This technique has been widely used by several news organizations such as the New York Times and The Guardian, and its effectiveness has been extensively studied \cite{bryan_temporal_2017, bradbury_documentary_2020, boy_storytelling_2015}. 
Segel and Heer \cite{segel_narrative_2010} highlighted the importance of balancing between author-driven parts of a narrative visualization, usually linear and non-interactive, and reader-driven ones, usually interactive and exploratory. Hullman and Diakopoulos \cite{hullman2015content} highlighted the importance of “rhetorical” and “persuasive” elements when communicating a story in visualizations. Drawing on both seminal papers, Lee et al. \cite{lee2015more} further articulated visual storytelling process starting from data exploration
to story delivery to users' feedback. }


\begin{figure}
    \centering
    \includegraphics[width=0.37\paperwidth]{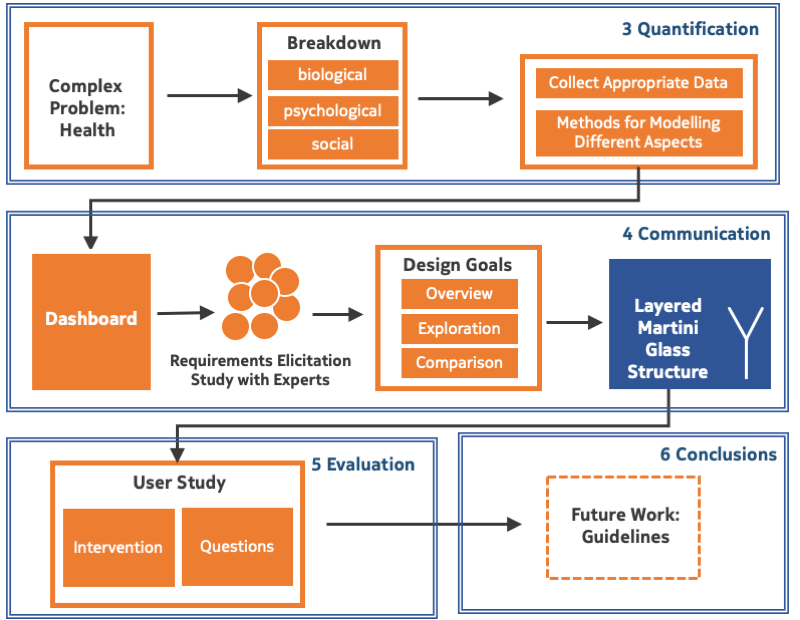}
    \caption{Steps to visualize the complexity behind health, and evaluate the resulting visualization.  After breaking down  the problem into its three main aspects, we quantified each of these aspects by collecting appropriate data upon which we run state-of-the-art machine learning models \emph{(top row)}.  We determined our design goals by conducting a requirements elicitation study with medical experts, and then built a layered Martini Glass Structure visualization \emph{(mid row)}. We conducted a user study upon which we drew guidelines for future reference \emph{(bottom row)}.}
    \label{fig:overview}
\end{figure}

In summary, previous work on \textit{quantifying} health outcomes has not been sufficiently general, in that, it has focused on very specific medical conditions  {or symptoms}, and it has done so by focusing on one condition/{symptom} at a time.  {Also, previous work on \emph{visually communicating} multi-faceted information in healthcare mainly targeted medical practitioners, and used dashboard-type of visualisations. 
}

\section{Quantifying the bio-psycho-social model}\label{sec:quantify}

We resorted to Reddit, an online community in which people exchange ideas and provide support to each other. Subcommunities in Reddit called subreddits consist of subsidiary threads that allow users to focus on a specific interest or topic. When deciding which subreddits to gather, we focused on the conditions with distinctive symptoms, making them easier to discriminate from others (e.g., psoriasis) and excluded complex and hard to discriminate conditions (e.g., cancer). 
Based on the data availability,  we incorporated the following $14$ conditions: rheumatoid arthritis, diabetes, depression, dementia, borderline personality disorder, psoriasis, gastroparesis, hypothyroidism, irritable bowel syndrome, dementia, Parkinson's disease, Meniere's disease, multiple sclerosis, and sleep apnea.

We  collated all the Reddit posts in those subreddits during the first half of 2017, which resulted in about half a million posts in total.

\subsection{Quantifying Biological Aspects}\label{sec:biodata}
The biological side of a medical condition is partly captured by symptoms and drugs associated with it. 

\paragraph{Expected Symptoms and Drugs from Literature.}
We used MedScape, WebMed, 
and Wikipedia. From MedScape, a website providing information to physicians and health care professionals, for a given condition, we obtained the list of signs and symptoms that are used to diagnose it. However, MedScape does not provide all the symptoms (i.e., it does not include the most common and generally non-discriminating symptoms, such as pain and inflammation). For this reason, we augmented this list of primary symptoms by manually searching WebMed and Wikipedia for additional ones. 

 Since the same symptom/drug can be expressed in several ways, doctors adopt a common technical terminology. This work adopted 
 the most comprehensive and widely used terminology, which is called the Systematized Nomenclature of Medicine -- Clinical Terms (SNOMED CT). It is used in over 50 countries, and contains synonym expressions for each medical concept (for example, for \textit{rheumatic arthritis}, synonyms include \textit{rheumatic gout} and \textit{proliferate arthritis}).  
We manually linked each of the previously collected symptoms/drugs to the corresponding concept in the SNOMED CT database. For instance, consider the symptom \textit{weakness}. Using SNOMED CT's search mechanism, we linked it to the concept \textit{asthenia} with the ID \textit{13791008}. As a result, we associated a set of \textit{expected symptoms and drugs} expressed in a proper medical language with each of the 14 medical conditions. For example, \emph{rheumatoid arthritis} was associated with the symptoms of \textit{stiffness, tenderness, swelling}, and \textit{pain in joints}, and with the drugs \textit{Leflunomide} and \textit{Celecoxib}, among others. 

\subsection{Quantifying Psychological Aspects}\label{quantifying-dictionary}
In addition to the biological aspects, a medical condition is characterized by its psychological aspects, i.e., a rich set of individual experiences.
To quantify such experiences, we needed to gather: i) \textit{all the symptoms/drugs mentioned by different patients} (a significantly larger set of symptoms/drugs emerges than the expected ones), as well as ii) the \textit{distinct emotions and body parts the patients mentioned}.

\noindent\paragraph{Emerging Symptoms and Drugs Online.} 
From the Reddit dataset, we extracted symptom and drug mentions. Extracting medical mentions from  {free-form} text is a challenging NLP task. People use different forms to express the same type of symptom (for instance, \textit{brain fog} and \textit{clouded consciousness}); use slang for medication names (for instance, \textit{benzos} for \textit{benzodiazepines}); or  make spelling mistakes. 
An equally challenging task is to map the extracted mentions to an official terminology, such as SNOMED CT. For example, one needs a method to link both \textit{brain fog} and \textit{clouded consciousness} to the same concept, which happens to have the ID \textit{40917007} in SNOMED CT. 

Thanks to the advances in modern deep-learning techniques in Natural Language Processing, it  has become possible to do both tasks. 
We implemented a state-of-the-art \emph{entity extraction method}  (sketched in Figure \ref{fig:models}, detailed in Supplemental Material, and published in  \cite{scepanovic2020medred}) based on bidirectional long short-term memory (BiLSTM) units in combination with conditional random fields (CRFs) that utilizes the contextual string embeddings (i.e., RoBERTa embeddings) \cite{liu2019roberta}. 



The symptoms/drugs mentioned on Reddit that were not among the expected ones were called \emph{emerging symptoms/drugs}. To see why, consider that, for \emph{rheumatoid arthritis}, in addition to its expected symptoms,
a myriad of additional symptoms had emerged online: \textit{hair loss, itching, exhaustion, bone pain, muscle weakness, mental dullness, heavy legs, tiredness,} and \textit{bedridden}, to name a few. Such emerging symptoms are not discriminative of the condition itself but, instead, they reveal the diversity of the individual experiences among patients, and part of them represent the psychological aspects of \textit{rheumatoid arthritis}.

\begin{figure}
    \centering
    \includegraphics[width=0.49\textwidth]{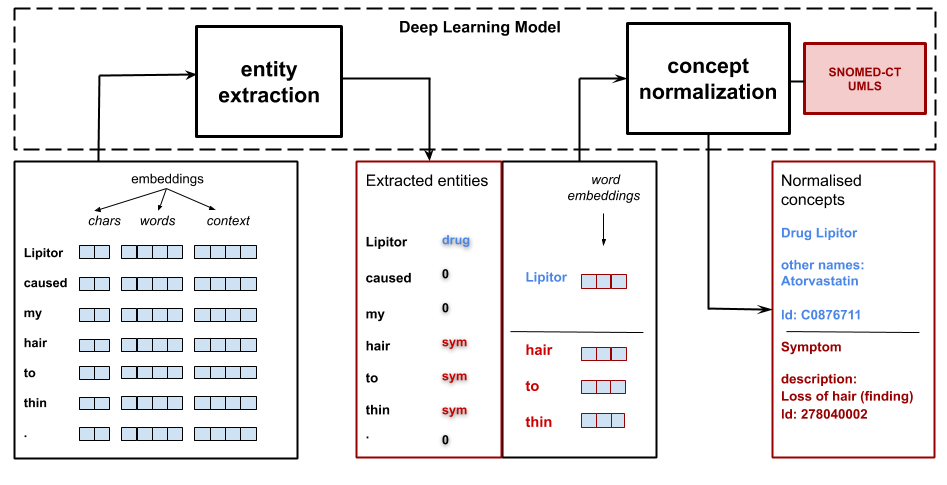}
    \caption{Sketch of our deep learning models that extract medical mentions from social media, and associate them with concepts present in our medical hierarchy (SNOMED CT).}
    \label{fig:models}
\end{figure}

\noindent\paragraph{Emotion Words.} 
One of the primary forms of people's psychological experience is represented by emotions. To categorize emotions extracted from our social media data, we employed the Plutchik's psychoevolutionary theory \cite{plutchik2001nature}. The theory lists 8 basic emotions: anger, joy, disgust, fear, anticipation, trust, sadness, and surprise. We looked at how many words expressing each of these basic emotions were mentioned when talking about each of the 14 conditions. To determine whether a word belonged to an emotion category, we resorted to the ``EmoLex'' word-emotion lexicon \cite{mohammad2010emotions}, which classifies words into the eight categories.

For each word (e.g.,~``cry'',``tear'') related to a specific emotion category (e.g.,~``sadness''),  and a condition (e.g.,~``depression''), we computed the Point-wise Mutual Information (PMI). Our PMI analysis (detailed in Supplemental Material) compares the observed occurrence probability to the chance probability of occurrence given a uni-gram model of independent interactions. It considers that words relevant to a keyword will occur with it far more often than would be expected by chance. 


By aggregating all the words (``sad'', ``tear'', ``cry'') belonging to a particular emotion category (``sadness''), we computed the association between each condition $d$ and each emotion category $cat$.

\paragraph{Body Parts.} 
There is previous research linking experience of emotions to different body parts \cite{nummenmaa2014bodily}, so our visualization showed which body parts people mentioned with each of the 14 conditions.
To determine the associations between body parts (e.g.,~ {``mouth''}) and a given medical conditions, we applied a methodology similar to the previous one used for emotions (i.e., a TF-IDF normalization). To that end, we created a `body' dictionary that lists all the words (e.g.,~``tongue'', ``lip'') related to a specific body part (e.g.,``mouth'') by combining the original LIWC dictionary subcategory ``body'' and the Wikipedia ``human body and organs.''\footnote{\url{https://en.wikipedia.org/wiki/List\_of\_organs\_of\_the\_human\_body}}

\subsection{Quantifying Social Aspects}\label{quantifying-prescriptions}
People's social environment can cause or affect their health. For example, the economic status has an influence on people's diets, and this, in turn, can affect whether they will suffer from diabetes \cite{aiello2019large}. On the other hand, people's medical condition can influence their social life -- for instance, cancer patients experiencing severe pain are likely to suffer from lowered-self esteem and anxiety \cite{ozdemiroglu2017self}. 
To associate socio-economic conditions with disease prevalence, we associated census data with prescription data across all the boroughs in England.

Prescription data is released by the National Health Service (NHS)
and enumerates all the prescriptions done by any general practice (GP) across England. 
The drug prescription data has been made  available since July 2010, is released every month by NHS~\cite{BNFsummary}, and consists of four files (Figure S2 the Supplemental Material), which we collected  until the end of June 2019:
i) \emph{NHS GP monthly prescriptions:} each prescription contains the drug's name, its  British National Formulary (BNF) code, the practice code, its total number of items, total cost, and each item's quantity; ii) \emph{NHS drugs:} each row contains a drug (individual preparation name) and its unique BNF Code; iii) \emph{NHS GPs:} each row contains a practice code, name, and full address; and iv) \emph{NHS GP patients:} each row contains a practice code, census LSOA borough code (each row refers to a specific (practice, LSOA) pair), and the number of the practice's patients in that borough. From this file, we  computed the total number of primary care patients who live in each borough.

The data, however, does not contain any mapping between the prescribed drugs and the corresponding medical conditions. To create such a mapping, we crawled the entire database from DrugBank\footnote{\url{drugbank.ca}}. From the database, we obtained $9105$ drug-related pages. Out of those, we were able to match $3013$ to at least one associated condition. This gave us a mapping between drugs and conditions. More specifically, for a given condition $C$, we used DrugBank to find the set of drugs $D$ associated with it to then estimate the total prescriptions related to $C$ in a borough $A$.

\section{Communicating the bio-psycho-social model}\label{sec:communicate}
We first developed a visualization for medical professionals, with whom we conducted a requirements elicitation study (Section \ref{sec:dashboard}). To fulfill the design goals that emerged from this study, we then surveyed and selected the most appropriate visualization approaches (Section \ref{sec:design_req}). These approaches were then used to develop the final visualization aimed at the general public (Section \ref{sec:martini}).

\subsection{Communicating with a Dashboard}\label{sec:dashboard}
 {The purpose of our study with medical professionals was twofold. First, we wanted them to assess the medical content and validity of the visualization. 
Second, as doctors convey medical knowledge to patients on a daily basis, we relied on their experience and asked them if the general public could be interested in such a visualization, and if so, what functionalities would be of help. From their answers, we selected several requirements and design suggestions.}

We found that the medical experts were already aware of the bio-psycho-social model, and
were familiar with dashboards \cite{bir_making_2019, concannon_developing_2019}.
Hence, for our elicitation study, we developed a dashboard prototype.

\begin{figure*}
    \centering
    \includegraphics[width=.88\linewidth]{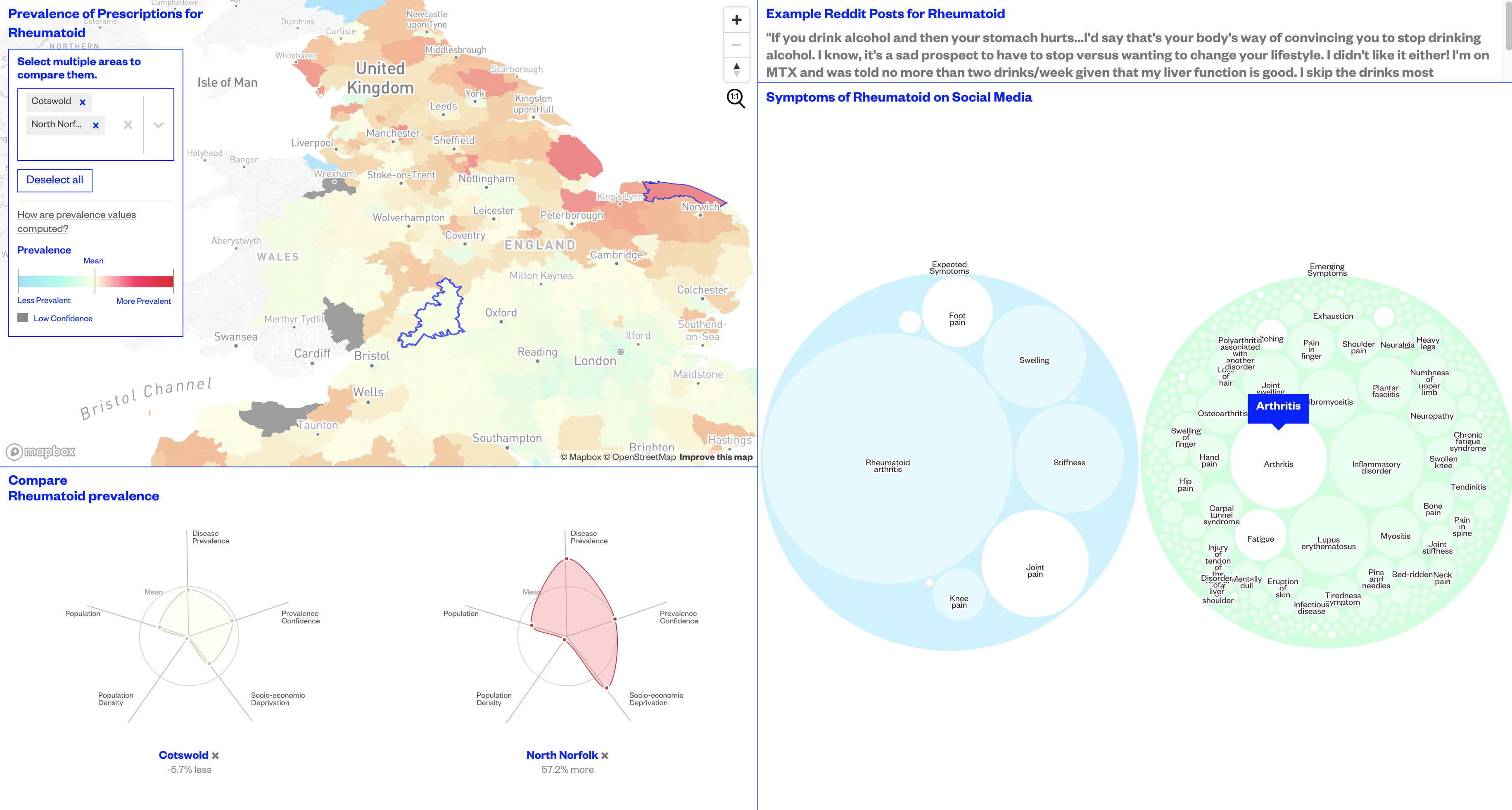}
    \caption{The dashboard used in the elicitation study among medical experts. A map encodes prescription prevalence together with socio-economic indicators \emph{(left)}, and blue/green bubbles show not only the typical symptoms but also those emerging from social media conversations \emph{(right)}.}
    \label{fig:dashboard_full}
\end{figure*}

Since dashboards aim at showing all the information at once, we provided the functionality to select a medical condition,
for which we then showed its real-world prevalence (left panel in Figure~\ref{fig:dashboard_full}) and associated symptoms (right panel in Figure~\ref{fig:dashboard_full}). When a user clicked on a region on the map, then the dashboard showed detailed information about the region, including the prevalence of the condition, the confidence value, and its socio-economic indicators such as deprivation index, population, and population density. The two sides of the dashboard were interactively linked; when a user clicked on a symptom bubble, then the choropleth map would change to display its prevalence. 

\emph{The goal} of the elicitation study
was to discover the main requirements for visualizing the bio-psycho-social model.

We recruited a group of 13 medical experts, who interacted with the dashboard and answered an online survey afterwards. Our domain experts were based in the United Kingdom (4 experts), United States (2), Italy (2) Ecuador (1), Kenya (1), Spain (1), and 2 preferred not to disclose their location; 3 of them were epidemiologists, 3 were Public Health researchers, 2 were clinical academics, 1 was a Public Health manager, while 4 preferred not to provide this information.

We asked them three open-ended questions:
\begin{enumerate}[noitemsep]
    \item[Q1] How would you describe the \emph{intuitiveness} of this tool? 
    \item[Q2] Could the general public be \emph{interested} in such a visualization and, if so, why? 
    \item[Q3] Which \emph{additional features} or functionalities would you like to see in a tool aimed at the general public?
\end{enumerate}

We then summarized their replies to each question.
\paragraph{[Q1] Intuitiveness.} 
In general, most of the experts found the tool to be intuitive. However, they also expressed difficulty with having the different aspects of a medical condition collated into a single screen. 
 They mentioned three main shortcomings: i) \emph{lack of self-explanation} (without a proper tutorial, it was hard to use the visualization to its full potential), ii) \emph{unclear links among data sources} (the respondents perceived that the combination of multiple datasets was not visually apparent, e.g., the relationship between the prescription map and the symptoms extracted), and
iii) \emph{screen real estate issue} (i.e., the space is too limited for communicating complex information).

\paragraph{[Q2] Interest of the General Public.} They found the tool could be of interest to general public for three main reasons: curiosity and novelty, a better understanding of the symptom-condition relationships, and ability to discover their areas' health.

In terms of \emph{curiosity} they said: \textit{``People might like to play around with such visualizations out of curiosity''}, \textit{``Absolutely - curiosity''}, \textit{``Yes. If done right, it can spark curiosity like (``oh I have depression, where are people similar to me?")''.} One expert pointed out the \emph{novelty} of the tool as a factor that could spark interest: \textit{``The general public would be interested for the novelty factor as there is no such tool that can bring these dimensions together.''}

Another aspect they mentioned was the ability to \emph{better understand the relationships between symptoms and medical conditions}.
For instance, professionals said that the tool \textit{``provides rapid information on symptoms associated with a specific disease and therapy"}, and \textit{``[...] it is a powerful tool to explore symptoms for lots of diseases.''} As \textit{``many diseases and conditions have symptoms that are uncommon and rare, doctors tend to dismiss these symptoms, leaving the patients confused and scared."} By using the tool, the general public can \textit{``feel how common is the disease and get a sense of companionship, if the set of signs and symptoms they present match a known disease.''} Finally, some of the experts said: \textit{``Also, it could be a good way to communicate news about health.''}, and \textit{``People are increasingly interested and wanting to be involved, share and contribute to health issues.''}

The third group of comments had to do with the \emph{health status of areas}. For instance, they mentioned: \textit{``Yes, because of the map: people would like to see what are the symptoms/diseases that are more common in the area they live in.''}; or \textit{``To have a general idea about diseases affecting their area.''}, or \textit{``I think they would be interested since they would like to know what the current disease outbreaks in neighborhoods so that they can prepare in advance.''}, and one expert went further to state \textit{``...figuring out the best places to live/visit, if the socio-economic indicators are incorporated.''}

\paragraph{[Q3] Additional Features.} 
The experts mentioned six main additional features: i) providing \emph{a tutorial} to guide the users through the complexity of the model and the visualization, ii) offering \emph{background} information about the results calculation, iii) highlighting \emph{uncommon symptoms}, iv) \emph{rankings of regions} according to the prevalence of medical conditions,  {v) \emph{improved labeling} of different parts of the visualization, and vi) \emph{adding mouse hovering interactions}.}

\paragraph{Design Goals.} After summarizing the requirements emerging from the elicitation study, we identified three main design goals:

\vspace{4pt}\noindent\textbf{Simplify user interaction and model exploration.} The final tool should guide interaction and exploration instead of asking users to make complex choices.

\vspace{4pt}\noindent\textbf{Support uncommon symptoms-condition exploration.} The tool should visually differentiate symptoms that are common to a condition from those that are uncommon.

\vspace{4pt}\noindent\textbf{Enable spatial exploration.} The tool should provide users with means of selecting several areas and comparing them in terms of different indicators.


\subsection{Communicating the Complexity}\label{sec:design_req}

To meet the three design goals, by  reviewing the literature, we adopted and identified three techniques: story synthesis, narrative visualization, and layering with animation.

First, we adopted \emph{story synthesis} \cite{chen2018supporting}. The idea  is to select and organize the quantitative findings.
Second, to show the organized findings, we adopted a \emph{narrative visualization} approach \cite{segel_narrative_2010}.
In particular, we adopted the \emph{Martini Glass Structure} identified by Segel and Heer \cite{segel_narrative_2010}, where the story begins with an explanation or narrative (``author-driven" part) and continues with a more exploratory and interactive data visualization (``reader-driven" part). Creating a linear-sequence would give us much more freedom than the dashboard because we can address each concept sequentially (“bio,” “psycho,” and “social”). Moreover, this Martini Glass-structured visualization could address the need of a tutorial without explicitly having it, as it can naturally increase the level of understanding while users go through the visualization.
 {{Third, we adopted the \emph{layering technique}  {\cite{dataurban_4_2019}}}} 
to reduce the complexity of the bio-psycho-social model with its multiple points of view.
{The layering technique} relies on breaking down the main point into multiple digestible chunks, and showing one chunk at a time. 
Also, drawing on Gestalt principles \cite{dataurban_4_2019}, several types of {animations} were incorporated (e.g., selection and scroll).
The effectiveness of animations lies in guiding the  {users' attention} to focus on the visualization's intended message. Such functionality allows the users 
to discover complex stories at their own pace.

To sum up, our design goals were met by a combination of three techniques: story synthesis, narrative visualization, and layering with animations. The \emph{Martini Glass structure} enabled a reader-driven approach (the user's selections determined what content was shown). 
To avoid complex user interactions, we incrementally introduced the three components of the bio-psycho-social model with \emph{layering} and \emph{scrolly-telling techniques}.

\subsection{Communicating with Layered Martini Glasses}\label{sec:martini}
\begin{figure}
    \centering
    \includegraphics[width=.777\linewidth]{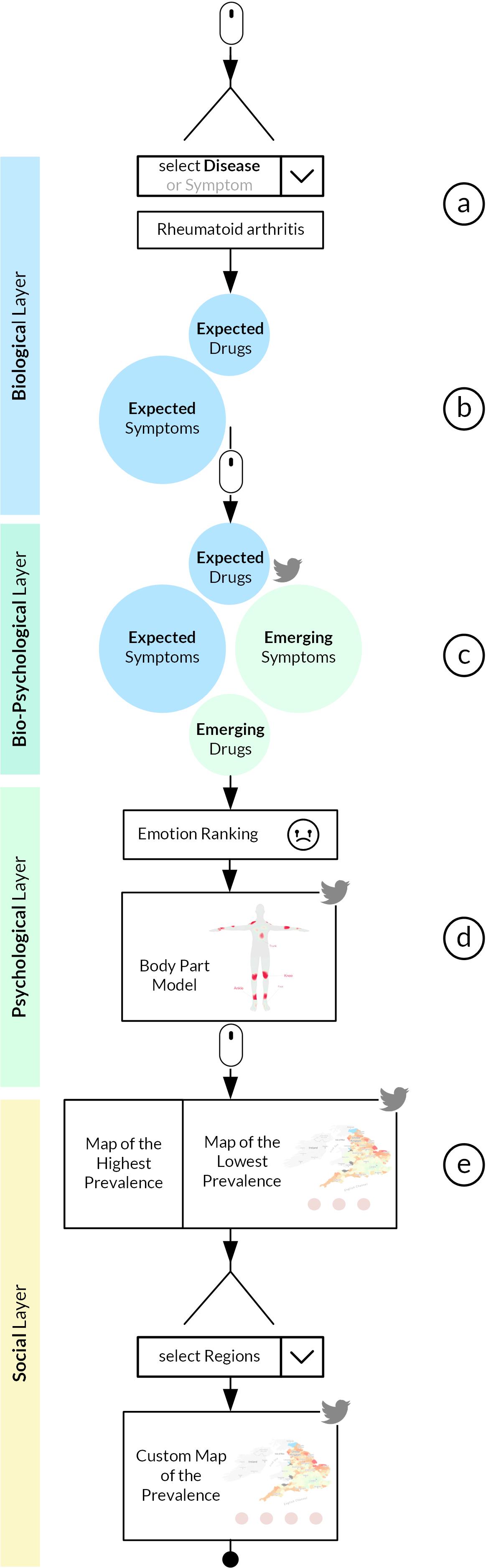}
    \caption{The storyline of the visualization. Upon selecting a medical condition (a), the three layers--the biological (b); (a view shown in Figure \ref{fig:psychoparts}, left), psychological (c and d); (Figure \ref{fig:psychoparts}, mid and right), and social (e); (Figure \ref{fig:regionsUK})--are linearly stacked with two Martini Glass structures (one in the biological layer, and the other in the social layer). Each layer offers opportunities to share parts of the visualization on social media.}
    \label{fig:glasses}
\end{figure}

We selected a suitable design representation for each of the three story slices: bubbles for the biological slice, human body and emoticons for the psychological, and a choropleth map for the social. Then we applied a layered Martini Glass structure (Figure~\ref{fig:glasses}) for the arrangement of the story slices.
The user interaction between the layers is provided by linear scrolly-telling and animations.
The first and last layers of the model are structured as two Martini Glasses. They start with our predefined author-driven narrative, and end up with a reader-driven exploration. Within each exploratory snippet, more complex interactions such as clicking, dragging, zooming, and panning are provided to offer personalized data views. The visualization has then multiple opportunities for users to share their 
``data views'' using social media.

\subsubsection{Introducing Biological Aspects} \label{viz-bio}
The visualization starts with a brief introduction of traditional biomedical model. The introduction, in turn, starts with the controversial statement that this  biomedical model ``doesn't fit the reality." This statement aims at capturing user's attention. Then, as a first interaction step, the  Martini Glass structure allows the user to select a medical condition (e.g., rheumatoid arthritis) (Figure \ref{fig:glasses}a).

Since traditional medical practice has focused on the biological side, the visualization starts with describing this layer by showing the \textit{expected} associated symptoms and drugs (Section \ref{sec:biodata}) for the selected condition, e.g., rheumatoid arthritis. The symptoms and drug names are represented as two blue Packed Bubble Charts (Figure \ref{fig:glasses}b and Figure \ref{fig:psychoparts}, left). The sizes of the bubbles vary based on how common the symptoms/drugs are, and is calculated using a TF-IDF normalization.  Once the user hovers over a bubble, the number and the list of all the associated conditions are displayed. 
 {We chose to use Packed Bubble Chart \cite{wang2006visualization} because of three main reasons. First, this type of diagram visualizes medical conditions as cohesive collections of symptoms/drugs with its aesthetically pleasing and almost organic appearance \cite{mylavarapu2019ranked}. Second, it visually encodes the frequency of symptoms/drugs with circle sizes that enable value comparisons \cite{heer2010tour}. Third, it enables easy discovery of unanticipated patterns of uncommon and overlapping symptoms/drugs.} 

We kept the blank space on the right hand side of the blue bubbles (Figure \ref{fig:glasses}b) to evoke tension and nudge the user to keep on scrolling down.

\subsubsection{Adding Psychological Aspects} \label{viz-psycho}
\begin{figure*}
    \centering
    \includegraphics[width=\linewidth]{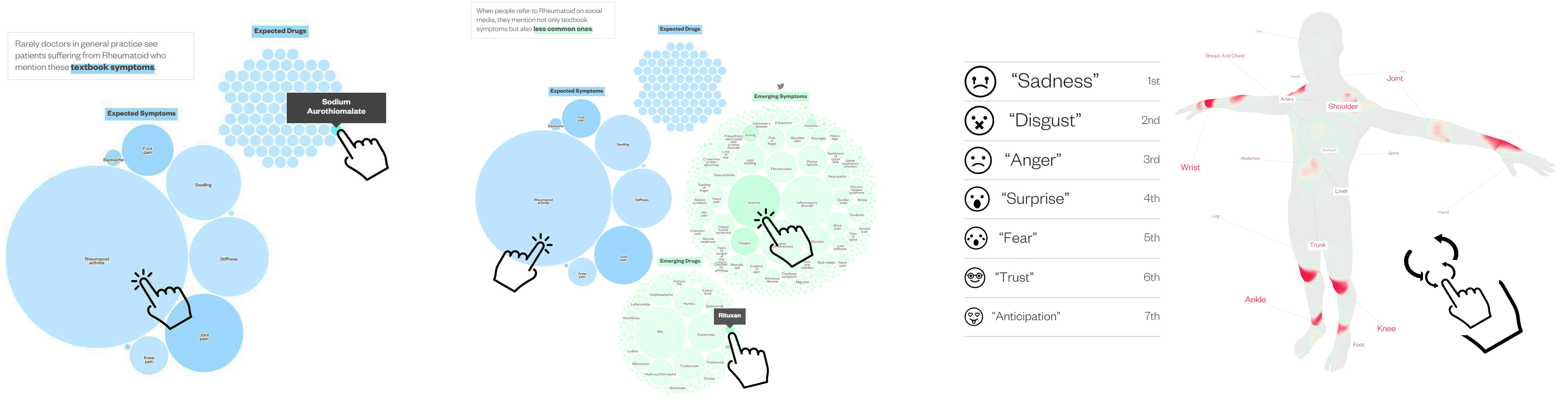}
    \caption{Visualizing the biological and psychological aspects. The packed circle charts display expected (left, blue bubbles) and emerging (middle, green bubbles) symptoms for \textit{rheumatoid arthritis}, together with corresponding drug names. The emotions associated with the condition and respective body parts are also shown (right).} \label{fig:psychoparts}
\end{figure*}

The  {visualization} of the second layer (Figure \ref{fig:glasses}c) starts with the green Packed Bubble Charts (Figure \ref{fig:psychoparts}, middle). The green clusters show the symptoms and drug names \textit{emerging} from social media. The animated transition from the expected symptoms and drug names to the emerging ones aims at softening the separation between these two types and at evoking surprise \cite{dataurban_4_2019}. By comparing these two bubble clusters, the user is able to acknowledge that social media data can uncover unforeseen symptoms and drug names. The 
``share button'' allows the user to post his/her current ``data view'' on Twitter.

Subsequently, the user is presented with emotions and associated body parts (Figure \ref{fig:glasses}d and Figure \ref{fig:psychoparts}, right). 
 {We tried to address the problem of intuitiveness previously emerged in relation to the dashboard and the need of showing emotional data with a direct visual encoding:  {(emoji-like)} icons  and a human body.} The first visualization introduces the ranking of the 7 emotions in terms of how often they are mentioned with rheumatoid arthritis, e.g., "sadness”, “disgust”, “anger”, “surprise”, “fear”, “trust,” and “anticipation.”  {We used emojis to represent the emotions  because of their common use \cite{danesi2016semiotics,ljubevsic2016global}.}
Below the emotion ranking, the user can see and rotate a 3D-model of a human body with a heat-map texture projected onto it. The texture illustrates specific body parts that social media users often mention when discussing the medical condition under consideration. We decided to mark this value with a texture on the surface of the body, instead of pointing at the exact body parts. This reduces the complexity of the visualization, and illustrates the inexact nature of social media expressions.  {The color and size of the body textures encode the magnitude of mentions. Our choice of including the body parts visualisation under the psychological aspects was motivated by previous work on bodily maps of emotions \cite{nummenmaa2014bodily}, which demonstrated that different emotions are associated with topographically recognizable bodily sensations.}
Again, the self-contained visualization can be fully shared by the user on Twitter.

\begin{figure}
    \centering
    \includegraphics[width=.88\linewidth]{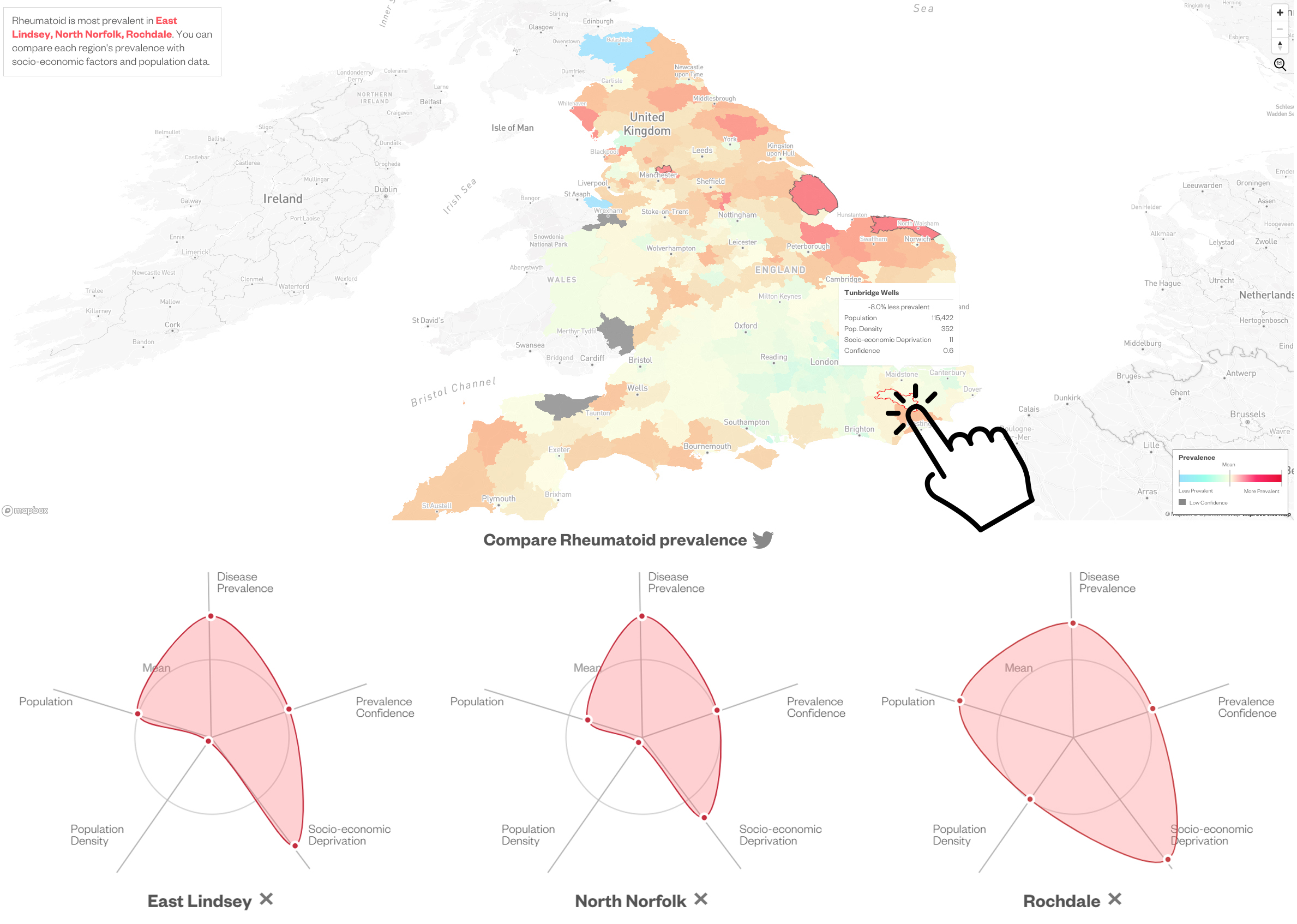}
    \caption{Visualizing the social aspects with:  a map of the prevalence of prescriptions for \textit{rheumatoid arthritis} \emph{(top)} (by clicking on a region, a pop-up window with its socio-economic indicators is shown); and  a comparison of the three regions with the highest prescription prevalence of the condition across England \emph{(bottom)}.}
    \label{fig:regionsUK}
\end{figure}

\subsubsection{Adding Social Aspects} \label{viz-socio}
The final layer (Figure \ref{fig:glasses}e) contains a choropleth map of England and a series of radar charts showing a region's population, population density, and socio-economic deprivation (Figure  \ref{fig:regionsUK}, top). The diverging color scheme shows the prevalence of the prescriptions associated with the  condition in relation to the mean value for England. The layer follows a Martini Glass structure. It first starts with an author-driven narrative, showing the regions with the highest and the lowest prevalence of prescriptions for the condition, and ends with  interactive maps. The user can select up to four regions by choosing them from the list, or by clicking them on the map. Then the radar charts are accordingly updated, so the user can compare the prevalences across the four regions (Figure \ref{fig:regionsUK}, bottom). For instance, the two regions with the highest rheumatoid arthritis prevalence are North Norfolk and Rochdale. The visualization shows that their socio-economic characteristics are almost the opposite: population size and population density of North Norfolk are way below the mean, while those of Rochdale are above the mean; and socio-economic deprivation is slightly above the mean for North Norfolk, while it is extremely high in Rochdale. { A radar chart, also known as star plot or polar chart \cite{chambers2018graphical,friendly2001milestones}, is a graphical method effective in visualizing multidimensional data with an arbitrary number of variables \cite{liu2008visualization,albo2015off,kim2012improved}. 
We selected the radar chart for two main reasons: \emph{i)} it allows for easy comparison of ordinal values of socio-economic indicators across regions, and \emph{ii)} it is suitable to convey outliers, i.e., the regions that have a value that is significantly above or below the country's average.} To further increase user engagement, again, the comparative maps can be shared on Twitter.

\section{Evaluation}\label{sec-evaluation}
The goal of the study was to evaluate how well our visualization communicated the complex bio-psycho-social model to an audience that might not be familiar with it. 

\subsection{User Study Setup}
The study consisted of four steps, taken from a method that was previously used for measuring the persuasive effects of data visualizations \cite{pandey_persuasive_viz_2014}. 

\vspace{4pt}\noindent\textbf{Step 1: Survey.} Once the participants agreed to take part in the experiment, we asked them to respond to a 6-statement survey that assessed their opinions on how doctors should operate.
Additionally, we asked them if they searched for health advices on the Internet and, if so, on which websites. Finally, we asked them to report their gender and age.

\vspace{4pt}\noindent\textbf{Step 2: Intervention.} We asked the participants to interact  {with our visualization (the treatment) or with the baseline (the control)} and to carefully investigate its content. This step relied on an attention-check question:  {we asked the treatment group to name two of the emotions associated with a medical condition and we asked the control group to name external websites the baseline visualization was referring to.}  
Answers to either question could be known only if the participants paid attention to the visualization.

\vspace{4pt}\noindent\textbf{Step 3: Survey.} After interacting with the  {treatment or control}, participants were asked again to respond to the survey. This allowed us to determine any potential opinion change due to the treatment/control.

\vspace{4pt}\noindent\textbf{Step 4: Open-ended questions.} We also asked the participants to elaborate on why they changed or did not change their opinion.  {For users interacting with the baseline, we also asked what charts or graphs should be added to the page to better explain symptoms of conditions, people's emotions, and their social circumstances. Participants could finally share any additional feedback.}

\paragraph{Survey.}
The survey consisted of six opinion determination statements. The first four statements started with the phrase \emph{``In treating medical conditions, doctors should..."} and ended as follows:

\vspace{4pt}\noindent\emph{... treat only the corresponding symptoms (S1)}. This statement aimed at assessing the acceptance level of the current biomedical model.

\noindent\emph{... consider also the emotional state of the patients (S2)}. This statement assessed the understanding and acceptance of the psychological aspects of conditions.

\noindent\emph{... consider also the social and economic backgrounds of the patients (S3)}. The statement assessed the understanding and acceptance of the social aspects of conditions.

\noindent\emph{... analyse what their patients shared on social media (S4)}. The statement revealed opinions towards using social media data in healthcare (impersonal perspective).

The last two statements were, instead, self-contained:

\noindent\emph{In treating your medical conditions, your doctor should analyse what you shared on social media (S5)}. This statement assessed whether the participants would share their own social media data for health purposes (personal perspective). 

\noindent\emph{Artificial Intelligence can improve healthcare (S6)}. The statement aimed at capturing their general opinions on the matter.

The participants were asked to consider each statement and decide to what extent they agreed with it on a 7-point Likert scale, ranging from ``strongly disagree" (-3) to ``strongly agree" (+3).

The experiment was conducted using the Amazon Mechanical Turk platform. The recruited participants fulfilled the following criteria: (i) self-reported location in the UK (to be able to relate to geographic prescription prevalence, which is shown on a UK map or  {described in the baseline}); (ii) previous task approval rate of at least $95 \%$; and (iii) number of previously approved tasks higher than $5000$. Each experiment took between 5-10 minutes, and participants were paid 1 USD for completing it.

\noindent\paragraph{Baseline.}
 The baseline for the control group (shown in Supplemental Material) had the following narrative structure: a short explanation of the selected condition extracted from Wikipedia; three pre-selected posts from Reddit mentioning symptoms, drug names, and emotions; and a table showing the prevalence of the condition and associated socio-economic factors. 

\subsection{Evaluation Metrics}
To measure potential opinion changes, one of the commonly used approaches is to segment the participants  into three \emph{opinion groups} based on their answers before and after exposure to the treatment/control \cite{pandey_persuasive_viz_2014}:
\begin{itemize}[leftmargin=*]
\item[] \emph{Negatively Polarized (NP)}: \emph{strongly disagree} (-3), and \emph{disagree} (-2);
\item[]\emph{Neutral/Weakly Polarized (NWP)}: \emph{somewhat disagree} (-1),  {\emph{neither agree nor disagree}} (0), and \emph{somewhat agree} (+1); 
\item[]\emph{Positively Polarized (PP)}: \emph{agree} (+2), and \emph{strongly agree} (+3).
\end{itemize}

By using these three groups, we calculated two metrics for each of the six statements. First, we calculated the \emph{persuasion likelihood} as the percentage growth rate of each opinion category ($\Delta NP$, $\Delta NWP$, and $\Delta PP$) after the user exposure to the visualization \cite{pandey_persuasive_viz_2014}. For example, the positively polarized category
$ \Delta PP= (PP_{after} - PP_{before}) $,
where $PP_{after}$ is the percentage of participants who were positively polarized towards a statement \emph{after} experiencing the visualization, and $PP_{before}$ is the percentage of participants who were so even \emph{before} experiencing it. In a similar way, we computed the persuasive likelihood across the two remaining opinion categories:
$ \Delta NWP= (NWP_{after} - NWP_{before}) $, and
$ \Delta NP= (NP_{after} - NP_{before}) $. Second, we  calculated the \emph{average opinion change} per statement, which is the mean across all participants of these $\Delta$s.

\subsection{Results}
 {The study was completed by 108 participants: 52 of them interacted with our visualization, and 56 with the baseline. The set of participants was diverse in terms of age (18-64 years, median age was 30), and gender (38\% females, 58\% males, and 4\% of the users preferred not to answer this question).}

\subsubsection{Opinions Before the Intervention (Step 1)}
Table \ref{table:1} summarizes the users' \emph{initial} opinions. The opinions on the current biological model were almost equally distributed among the different groups. Considering the three aspects of the bio-psycho-social model, the most positive opinions were towards its psychological aspect: 63\% of all participants agreed that doctors should consider the emotional state of their patients. The idea of including social and economic backgrounds tended to elicit either neutral responses (44\%) or positive ones (42\%). Unsurprisingly, participants were considerably against the general idea of sharing social media data with doctors (65\%). This opinion got even more polarized once they were asked to think about their own data (71\% of negative answers). The majority of the participants already had a neutral (46\%) or positive opinion (50\%) towards the use of AI in healthcare.

To search for reliable health advises our participants usually browsed professional medical websites such as the official portal of NHS (40 mentions) and WebMD (17 mentions). The trend of using social media as a source of health-related information was weak among them -- only 3 mentioned the use of Reddit, but 24 said they performed general health searches on Google.

\begin{table}[h!]
\centering
	\small
\begin{tabular}{ m{0.2 cm} m{5cm} m{0.38cm} m{0.38cm}  m{0.38cm} } 
 \hline
 $\#$ & Statement  & NP & NWP & PP \\
 \hline
 \rowcolor{Gray} \multicolumn{5}{l}{In treating medical conditions, doctors should:} \\
 S1 & treat only the corresponding symptoms & .31 & \textbf{.37} & .33 \\ 
 S2 & consider also the emotional state of patients  & .04 & .33 & \textbf{.63} \\
 S3 & consider also the socio-economic background of patients & .13 & \textbf{.44} & .42 \\
 S4 & analyse what their patients shared on social media & \textbf{.65} & .23 & .12 \\
 \rowcolor{Gray}  \multicolumn{5}{l}{In treating your medial conditions, your doctor should:} \\
 S5 & analyse what you shared on social media & \textbf{.71} & .19 & .10 \\
 \rowcolor{Gray}  \multicolumn{5}{l}{Artificial Intelligence can:} \\
 S6 & improve healthcare & .04 & .46 & \textbf{.50} \\
 \hline
\end{tabular}
\vspace{0.1cm}
\caption{Distribution of the participants' initial opinions  per statement. The percentages of responses in each of the negatively polarized (NP), negatively/weakly polarized (NWP), or positively polarized (PP) categories are shown. Participants' opinions were  equally distributed across the categories for the biological aspect  (S1), tended to be positive about the psychological aspect (S2), and were split between positive and neutral views about the sociological aspect (S3). When it came to the use of social media data in healthcare, our participants were  negatively polarized (S4 and S5). The  use of Artificial Intelligence was mostly viewed as being either positive or neutral.}
\label{table:1}
\end{table}

\subsubsection{Opinions After the Intervention (Step 3)}

To test for statistical significance of the observed patterns, we used the Wilcoxon Signed-Rank Test, which tested the null hypothesis that the exposure to the  treatment or control did not affect a participant's opinion. The findings were considered statistically significant at the level of $p < .05$.

\paragraph{Persuasion Likelihood.} Table \ref{table:results} shows the persuasion likelihood by statement  after being exposed to our visualization. The participants tended to slightly decrease their support for the biological model (S1). Those who were neutral about the importance of psychological aspects switched their opinion to a more positive view ($\Delta$ 10\%) (S2), while those neutral about the importance of socio-economic conditions took either a more negative or more positive stance and, as such, became more polarized (S3). 

Those who were against the use of social media switched to stances that were more neutral (21\%) or more positive (6\%) (S4 and S5). Finally, 10\% switched to a more positive view of the general role of Artificial Intelligence in healthcare (S6).

\begin{table}[h!]
	\centering
	\small
	\begin{tabular}{ m{0.2 cm} m{5cm} m{0.4cm} m{0.4cm}  m{0.4cm} } 
		\hline
		$\#$ & Statement & NP & NWP & PP \\
		\hline
		\rowcolor{Gray} \multicolumn{5}{l}{In treating medical conditions, doctors should:} \\
		S1 & treat only the corresponding symptoms & \cellcolor{blue!10}  .00 & \cellcolor{green!10} \textbf{.04}  & \cellcolor{magenta!10} -.04 \\ 
		S2 & consider also the emotional state of patients  & \cellcolor{blue!10} .00 & \cellcolor{magenta!10} -.10 & \cellcolor{green!10} \textbf{.10} \\
		S3 & consider also the socio-economic background of patients &\cellcolor{green!10}  .06 & \cellcolor{magenta!10}  -.12 & \cellcolor{green!10} .06 \\
		S4 & analyse what their patients shared on social media & \cellcolor{magenta!10}  -.27 & \cellcolor{green!10} \textbf{.21} & \cellcolor{green!10} .06 \\
		\rowcolor{Gray}  \multicolumn{5}{l}{In treating your medial conditions, your doctor should:} \\
		S5 & analyse what you shared on social media & \cellcolor{magenta!10} -.27 & \cellcolor{green!10}  \textbf{.21} & \cellcolor{green!10}  .06 \\
		\rowcolor{Gray}  \multicolumn{5}{l}{Artificial Intelligence can:} \\
		S6 & improve healthcare & \cellcolor{magenta!10}  -.04 & \cellcolor{magenta!10}  -.06 & \cellcolor{green!10} \textbf{.10} \\
		\hline
	\end{tabular}
	\vspace{0.1cm}
	\caption{Persuasion likelihood  -- the difference in opinions before and after interacting with {our} visualization. Color scheme: no change (blue), negative change (magenta), positive change (green). Participants changed their opinions on the emotional aspects of conditions (S2), the use of social media data (S4 and S5), and on the use of AI (S6) in healthcare.}
	\label{table:results}
\end{table}

\begin{figure}
    \centering
    \includegraphics[width=0.9\linewidth]{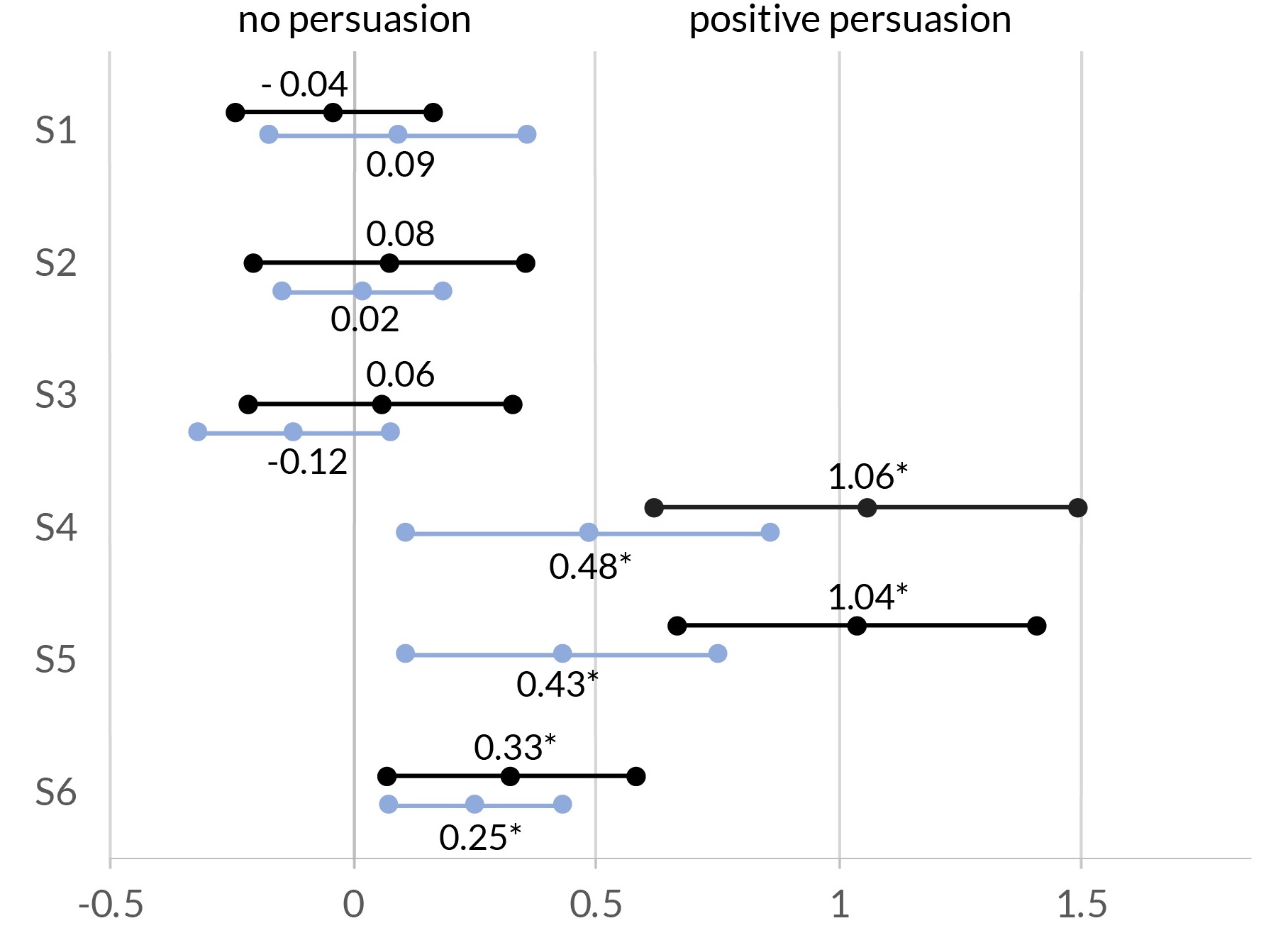}
    \caption{Statement-wise average opinion change with a 95\% confidence. {The values for those exposed to our visualization are marked in black, and for those exposed to the baseline in blue.} The opinion changes on the three last statements were statistically significant and, as such, are marked with a *.}
    \label{fig:persuasion}
\end{figure}

\paragraph{Average Opinion Change.} 
Figure \ref{fig:persuasion} shows the average opinion change by statement at a 95\% confidence. Changes were statistically significant for the last three statements  {for both baseline and our visualization}. The highest changes concerned the sharing of the social media data with doctors to improve treatment possibilities (S4 and S5).  {Overall, the persuasive effect of the visualization was stronger than that of the baseline.}

\subsubsection{Answers to the Open-ended Questions (Step 4)}
This section summarizes the reasons why the participants decided to change or not their opinions.

To begin with, to test whether our study with domain experts correctly identified the needs of the general public, we asked the baseline participants (those being exposed to the control) how the bio-psycho-social model should be visualized. The answers of both groups turned out to be aligned. Purely based on reading the Wikipedia and Reddit posts, baseline participants mentioned they would like to see a \textit{``list of symptoms belonging to each subreddit, automatically derived from mining posts''} and \textit{``how common symptoms are.''} Regarding the psychological side, the baseline participants suggested to show \textit{``the emotions of people in relation to their diagnosis.''} Finally, they suggested the social side could be presented as \textit{``a geographical map of the prevalence of diseases.''}


By then comparing the baseline participants with the treatment participants, we found three emerging themes.

\paragraph{Bio-psycho-social Model.}  {
The most convincing visual elements were the \textit{body part visualization} (4 mentions), the \textit{bubbles of symptoms and diseases} (4 mentions), and the \textit{radar charts} with socio-economic conditions (2 mentions).  {Similarly, in the baseline group, participants were also influenced by the emotions expressed in the posts (5 mentions), yet, the symptoms (2 mentions), and socio-economic conditions (1 mention) were less important to them.}}

The main reasons for not changing opinion, on the other hand, were two: the participants either already agreed with the benefits of the bio-psycho-social model (\textit{``I already felt this way and the visualization solidified this.''}), or trusted the dominant medical model and, as such, were reluctant to change (\textit{``Bio-medical model which doctors use is very trustworthy, or at least gives this image.''}).

\paragraph{Social Media Data for Healthcare.} In relation to social media use, the reasons for adopting a more favourable view included the presence of the \textit{body-part model and emotion ranking} (4 mentions), and of the \textit{bubbles of symptoms and conditions} (2 mentions). One participant was positively impressed by \textit{the variety of social media data} used in the project. The interaction with the psychological layer had a particular effect on our participants (\textit{``it made sense that social media can be considered a record of mental state over time''}, and \textit{``people struggle when put in a position where they have to verbalise their feelings to a doctor, they seem to talk more freely on social media these days.''}). The reasons for not changing opinion, on the other hand, were two: the participants were skeptical towards data sources ( \textit{``It just makes sense and could be very helpful if the data collected is real.''}), or they were concerned about privacy issues (\textit{``Although in my opinion I moved to be less against using social media I still feel the privacy issues are more important.''}).  {The baseline participants mentioned  privacy issues more often than the participants exposed to our visualization, likely because the former were exposed to the raw data.}

\paragraph{AI for Healthcare.} In relation to AI use, the reasons for changing opinions included the visualization elements, such as the \textit{map in combination with the radar chart} (4 mentions), the \textit{added value of merging social media data with official data sources} (3 users), the \textit{bubbles of symptoms and diseases} (2 users), the \textit{body part visualization} (2 users), and the \textit{overall story} flow (1 user). The reasons for not changing opinion, on the other hand, were that the participants had been exposed to the benefits of AI for healthcare before (\textit{``This technology [AI] is already a huge part of the healthcare.''}), or they required a proof of reliability of AI in healthcare (\textit{``this part of AI is still to be checked if it works 100\%.''}).

\section{Discussion and Conclusion}
We provided a way of {visually communicating} complex and multi-faceted health information and found that our approach based on a layered narrative was effective not only in creating user awareness but also in promoting opinion change. Our results speak to the importance of visualizations in research fields beyond InfoVis. Take Data Science, for example. This field uses the very same AI methods and systems presented in this paper, yet often neglects information visualization aspects.
Yet, these aspects are increasingly becoming critical, from introducing ``humans in the loop'' in the production of training data, to presenting model predictions in adaptive ways. 

\subsection{Limitations}
 {
Our work comes with two main limitations:
\begin{description}
\item \vspace{2pt}\noindent\textit{Biases of social media data.} While social media data has proven to be a rich and valuable source of health-related information, it carries inherent biases in terms of the non-representative user bases, 
self-selection biases, 
and the existence of bots (i.e., non-human users). 
\item \vspace{2pt}\noindent\textit{Testing a single type of visual storytelling.} We have evaluated our Layered Martini Glass visualization against a text-based baseline. However, other types of visual approaches could have been adopted.
\end{description}
}

\subsection{Theoretical implications}
We identified two main  {theoretical} aspects that the community could continue to explore in the future:
\begin{description}
\item \vspace{2pt}\noindent\textit{Adaptive narrative visualization.} Health knowledge is distributed differently across different people and roles. Depending on a user's knowledge, a future visualization could automatically adapt the narrative it shows (e.g., a street-level person could be exposed to a different narrative than that of a doctor).
\item \vspace{2pt}\noindent\textit{ {Visual communication} with human-in-the-loop.} Training data (e.g., social media data) is inherently biased. One way of partly tackling that problem is to introduce ``humans in the loop'' in the form of, for example, \emph{i)} professionals flagging any incorrect input data or AI results; and \emph{ii)} patients providing training data based on their personal health experiences.
\end{description}

\subsection{Practical implications}
 {
We identified two main practical implications of our work:
\begin{description}
\item \vspace{2pt}\noindent\textit{Enhancing healthcare communication.} The participants of the study saw our visualization as a new medium of communication between doctors and patients: \textit{``doctors will have a reference in each area to ask about other symptoms and problems.''}, 
and \textit{``as people seem to talk more freely on social media these days''} the verbalisation of feelings to the doctors could \textit{``fast track their diagnosis.''}
\item \vspace{2pt}\noindent\textit{Visual functionalities for social media platforms.} The participants found the visual storytelling superior to the text-based storytelling, which mimicked social-media platform interfaces. This finding suggests that social media platforms could add new functionalities similar to our visualization tool.
\end{description}
}

\acknowledgments{
The authors thank Enrique Martin-Lopez who had helped to analyze the social media data.
}

\balance

\bibliographystyle{abbrv-doi}

\bibliography{healthviz}

\begin{thebibliography}{10}

\bibitem{BNFsummary}
{BNF Classifications}.
\newblock
  \url{https://digital.nhs.uk/data-and-information/areas-of-interest/prescribing/practice-level-prescribing-in-england-a-summary},
  2019.
\newblock [Online; accessed Oct. 2019].

\bibitem{aiello2019large}
L.~M. Aiello, R.~Schifanella, D.~Quercia, and L.~Del~Prete.
\newblock Large-scale and high-resolution analysis of food purchases and health
  outcomes.
\newblock {\em EPJ Data Science}, 8(1):14, 2019.

\bibitem{albo2015off}
Y.~Albo, J.~Lanir, P.~Bak, and S.~Rafaeli.
\newblock Off the radar: Comparative evaluation of radial visualization
  solutions for composite indicators.
\newblock {\em IEEE Transactions on Visualization and Computer Graphics},
  22(1):569--578, 2015.

\bibitem{bayat2011symptoms}
N.~Bayat, G.~H. Alishiri, A.~Salimzadeh, M.~Izadi, D.~K. Saleh, M.~M.
  Lankarani, and S.~Assari.
\newblock Symptoms of anxiety and depression: A comparison among patients with
  different chronic conditions.
\newblock {\em Journal of research in medical sciences: the official journal of
  Isfahan University of Medical Sciences}, 16(11):1441, 2011.

\bibitem{bir_making_2019}
A.~Bir, N.~Freeman, R.~Chew, K.~Smith, J.~Derzon, and T.~Day.
\newblock {Making Evidence Actionable: Interactive Dashboards, Bayes, and
  Health Care Innovation}.
\newblock {\em {eGEMs: The Journal of Electronic Health Data and Methods}},
  7(1):40, Aug. 2019.

\bibitem{boy_storytelling_2015}
J.~Boy, F.~Detienne, and J.-D. Fekete.
\newblock Storytelling in information visualizations: Does it engage users to
  explore data?
\newblock In {\em Proceedings of the ACM Conference on Human Factors in
  Computing Systems}, pp. 1449--1458, Apr. 2015.

\bibitem{bradbury_documentary_2020}
J.~D. Bradbury and R.~E. Guadagno.
\newblock Documentary narrative visualization: Features and modes of
  documentary film in narrative visualization.
\newblock {\em Information Visualization}, 19(4):339--352, 2020.

\bibitem{briant2016power}
K.~J. Briant, A.~Halter, N.~Marchello, M.~Escare{\~n}o, and B.~Thompson.
\newblock The power of digital storytelling as a culturally relevant health
  promotion tool.
\newblock {\em Health promotion practice}, 17(6):793--801, 2016.

\bibitem{bryan_temporal_2017}
C.~Bryan, K.-L. Ma, and J.~Woodring.
\newblock Temporal summary images: An approach to narrative visualization via
  interactive annotation generation and placement.
\newblock {\em {IEEE Transactions on Visualization and Computer Graphics}},
  23(1):511--520, 2016.

\bibitem{chambers2018graphical}
J.~M. Chambers.
\newblock {\em Graphical methods for data analysis}.
\newblock CRC Press, 2018.

\bibitem{chen2018supporting}
S.~{Chen}, J.~{Li}, G.~{Andrienko}, N.~{Andrienko}, Y.~{Wang}, P.~H. {Nguyen},
  and C.~{Turkay}.
\newblock Supporting story synthesis: Bridging the gap between visual analytics
  and storytelling.
\newblock {\em IEEE Transactions on Visualization and Computer Graphics},
  26(7):2499--2516, July 2020.

\bibitem{de2014mental}
M.~D. Choudhury and S.~De.
\newblock {Mental Health Discourse on Reddit: Self-Disclosure, Social Support,
  and Anonymity}.
\newblock In {\em Proceedings of the International AAAI Conference on Weblogs
  and Social Media}, May 2014.

\bibitem{cohen2005survey}
A.~M. Cohen and W.~R. Hersh.
\newblock A survey of current work in biomedical text mining.
\newblock {\em Briefings in bioinformatics}, 6(1):57--71, 2005.

\bibitem{concannon_developing_2019}
D.~Concannon, K.~Herbst, and E.~Manley.
\newblock Developing a {Data} {Dashboard} {Framework} for {Population} {Health}
  {Surveillance}: {Widening} {Access} to {Clinical} {Trial} {Findings}.
\newblock {\em JMIR formative research}, 3(2):e11342, Apr. 2019.

\bibitem{consedine2007role}
N.~S. Consedine and J.~T. Moskowitz.
\newblock The role of discrete emotions in health outcomes: A critical review.
\newblock {\em Applied and Preventive Psychology}, 12(2):59--75, 2007.

\bibitem{Culotta204twitter}
A.~Culotta.
\newblock {Estimating County Health Statistics with Twitter}.
\newblock In {\em Proceedings of the ACM Conference on Human Factors in
  Computing Systems}, pp. 1335--1344, Apr. 2014.

\bibitem{curtis2019opioid}
H.~J. Curtis, R.~Croker, A.~J. Walker, G.~C. Richards, J.~Quinlan, and
  B.~Goldacre.
\newblock Opioid prescribing trends and geographical variation in england,
  1998--2018: a retrospective database study.
\newblock {\em The Lancet Psychiatry}, 6(2):140--150, 2019.

\bibitem{cutler2008socioeconomic}
D.~M. Cutler, A.~Lleras-Muney, and T.~Vogl.
\newblock Socioeconomic status and health: dimensions and mechanisms.
\newblock Technical report, National Bureau of Economic Research, 2008.

\bibitem{danesi2016semiotics}
M.~Danesi.
\newblock {\em The semiotics of emoji: The rise of visual language in the age
  of the internet}.
\newblock Bloomsbury Publishing, 2016.

\bibitem{elias_exploration_2011}
M.~Elias and A.~Bezerianos.
\newblock Exploration {Views}: {Understanding} {Dashboard} {Creation} and
  {Customization} for {Visualization} {Novices}.
\newblock In {\em Human-{Computer} {Interaction} – {INTERACT}}. Springer
  Berlin Heidelberg.

\bibitem{engel_need_1977}
G.~L. Engel.
\newblock The need for a new medical model: a challenge for biomedicine.
\newblock {\em Science}, 196(4286):129--136, 1977.

\bibitem{faiola_supporting_2015}
A.~Faiola, P.~Srinivas, and J.~Duke.
\newblock Supporting {Clinical} {Cognition}: {A} {Human}-{Centered} {Approach}
  to a {Novel} {ICU} {Information} {Visualization} {Dashboard}.
\newblock {\em AMIA Annual Symposium Proceedings}, 2015:560--569, Nov. 2015.

\bibitem{fava2007biopsychosocial}
G.~A. Fava and N.~Sonino.
\newblock The biopsychosocial model thirty years later.
\newblock {\em Psychotherapy and psychosomatics}, 77(1):1, 2007.

\bibitem{friendly2001milestones}
M.~Friendly and D.~J. Denis.
\newblock Milestones in the history of thematic cartography, statistical
  graphics, and data visualization.
\newblock \url{http://www.datavis.ca/milestones/}, 2001.
\newblock [Online; accessed Oct. 2019].

\bibitem{gonzalez2017capturing}
G.~Gonzalez-Hernandez, A.~Sarker, K.~O'Connor, and G.~Savova.
\newblock Capturing the patient's perspective: a review of advances in natural
  language processing of health-related text.
\newblock {\em Yearbook of Medical Informatics}, 26(01):214--227, 2017.

\bibitem{heer2010tour}
J.~Heer, M.~Bostock, and V.~Ogievetsky.
\newblock A tour through the visualization zoo.
\newblock {\em Communications of the ACM}, 53(6):59--67, 2010.

\bibitem{hullman2015content}
J.~Hullman, N.~Diakopoulos, E.~Momeni, and E.~Adar.
\newblock {Content, context, and critique: Commenting on a data visualization
  blog}.
\newblock In {\em Proceedings of the ACM conference on Computer Supported
  Cooperative Work \& Social Computing}, pp. 1170--1175, Feb. 2015.

\bibitem{big-data-engagement}
H.~{Kennedy}, R.~L. {Hill}, W.~{Allen}, and A.~{Kirk}.
\newblock Engaging with (big) data visualizations: Factors that affect
  engagement and resulting new definitions of effectiveness.
\newblock {\em First Monday}, 21(11), 2016.

\bibitem{khairat2018impact}
S.~S. Khairat, A.~Dukkipati, H.~A. Lauria, T.~Bice, D.~Travers, and S.~S.
  Carson.
\newblock The impact of visualization dashboards on quality of care and
  clinician satisfaction: integrative literature review.
\newblock {\em JMIR human factors}, 5(2):e22, 2018.

\bibitem{kim2012improved}
J.~H. Kim, V.~Iyer, S.~B. Joshi, D.~B. Volkin, and C.~R. Middaugh.
\newblock Improved data visualization techniques for analyzing macromolecule
  structural changes.
\newblock {\em Protein Science}, 21(10):1540--1553, 2012.

\bibitem{lanchantin_deep_2016}
J.~Lanchantin, R.~Singh, B.~Wang, and Y.~Qi.
\newblock Deep motif dashboard: visualizing and understanding genomic sequences
  using deep neural networks.
\newblock {\em Pacific Symposium on Biocomputing}, Nov. 2016.

\bibitem{leaman2010towards}
R.~Leaman, L.~Wojtulewicz, R.~Sullivan, A.~Skariah, J.~Yang, and G.~Gonzalez.
\newblock Towards internet-age pharmacovigilance: extracting adverse drug
  reactions from user posts to health-related social networks.
\newblock In {\em Proceedings of the Worshop on Biomedical Natural Language
  Processing}, pp. 117--125. Association for Computational Linguistics, 2010.

\bibitem{lee2015more}
B.~Lee, N.~H. Riche, P.~Isenberg, and S.~Carpendale.
\newblock More than telling a story: Transforming data into visually shared
  stories.
\newblock {\em IEEE Computer Graphics and Applications}, 35(5):84--90, 2015.

\bibitem{lee2015enhanced}
H.-C. Lee, Y.-Y. Hsu, and H.-Y. Kao.
\newblock An enhanced crf-based system for disease name entity recognition and
  normalization on biocreative v dner task.
\newblock In {\em Proceedings of the BioCreative Challenge Evaluation
  Workshop}, pp. 226--233, 2015.

\bibitem{big-data-health}
R.~{Lin}, Z.~{Ye}, H.~{Wang}, and B.~{Wu}.
\newblock Chronic diseases and health monitoring big data: A survey.
\newblock {\em IEEE Reviews in Biomedical Engineering}, 11:275--288, 2018.

\bibitem{big-data-viz-advances}
S.~{Liu}, D.~{Maljovec}, B.~{Wang}, P.~{Bremer}, and V.~{Pascucci}.
\newblock Visualizing high-dimensional data: Advances in the past decade.
\newblock {\em IEEE Transactions on Visualization and Computer Graphics},
  23(3):1249--1268, 2017.

\bibitem{liu2008visualization}
W.-Y. Liu, B.-W. Wang, J.-X. Yu, F.~Li, S.-X. Wang, and W.-X. Hong.
\newblock Visualization classification method of multi-dimensional data based
  on radar chart mapping.
\newblock In {\em International Conference on Machine Learning and
  Cybernetics}, vol.~2, pp. 857--862. IEEE, 2008.

\bibitem{liu2019roberta}
Y.~Liu, M.~Ott, N.~Goyal, J.~Du, M.~Joshi, D.~Chen, O.~Levy, M.~Lewis,
  L.~Zettlemoyer, and V.~Stoyanov.
\newblock {RoBERTa: A robustly optimized bert pretraining approach}.
\newblock {\em arXiv preprint arXiv:1907.11692}, 2019.

\bibitem{ljubevsic2016global}
N.~Ljube{\v{s}}i{\'c} and D.~Fi{\v{s}}er.
\newblock A global analysis of emoji usage.
\newblock In {\em Proceedings of the Web as Corpus Workshop}, pp. 82--89, 2016.

\bibitem{mccurdy2016visual}
K.~McCurdy.
\newblock Visual storytelling in healthcare: Why we should help patients
  visualize their health.
\newblock {\em Information Visualization}, 15(2):173--178, 2016.

\bibitem{mohammad2010emotions}
S.~M. Mohammad and P.~D. Turney.
\newblock Emotions evoked by common words and phrases: Using mechanical turk to
  create an emotion lexicon.
\newblock In {\em Proceedings of the NAACL HLT Workshop on Computational
  Approaches to Analysis and Generation of Emotion in Text}, pp. 26--34.
  Association for Computational Linguistics, 2010.

\bibitem{mylavarapu2019ranked}
P.~Mylavarapu, A.~Yalcin, X.~Gregg, and N.~Elmqvist.
\newblock Ranked-list visualization: A graphical perception study.
\newblock In {\em Proceedings of the Conference on Human Factors in Computing
  Systems}, pp. 1--12, 2019.

\bibitem{nikfarjam2015pharmacovigilance}
A.~Nikfarjam, A.~Sarker, K.~O'Connor, R.~Ginn, and G.~Gonzalez.
\newblock Pharmacovigilance from social media: mining adverse drug reaction
  mentions using sequence labeling with word embedding cluster features.
\newblock {\em Journal of the American Medical Informatics Association},
  22(3):671--681, 2015.

\bibitem{nummenmaa2014bodily}
L.~Nummenmaa, E.~Glerean, R.~Hari, and J.~K. Hietanen.
\newblock Bodily maps of emotions.
\newblock {\em Proceedings of the National Academy of Sciences},
  111(2):646--651, 2014.

\bibitem{ozdemiroglu2017self}
F.~Ozdemiroglu, C.~O. Memis, N.~Meydan, B.~Dogan, S.~M. Kilic, L.~Sevincok, and
  K.~Karakus.
\newblock Self-esteem, pain and suicidal thoughts in a sample of cancer
  patients.
\newblock {\em Psychiatry and Behavioral Sciences}, 7(3):156, 2017.

\bibitem{pandey_persuasive_viz_2014}
A.~V. {Pandey}, A.~{Manivannan}, O.~{Nov}, M.~{Satterthwaite}, and
  E.~{Bertini}.
\newblock The persuasive power of data visualization.
\newblock {\em IEEE Transactions on Visualization and Computer Graphics},
  20(12):2211--2220, 2014.

\bibitem{park2017tracking}
A.~Park and M.~Conway.
\newblock Tracking health related discussions on reddit for public health
  applications.
\newblock In {\em AMIA Annual Symposium Proceedings}, vol. 2017, p. 1362.
  American Medical Informatics Association, 2017.

\bibitem{park2018examining}
A.~Park, M.~Conway, and A.~T. Chen.
\newblock Examining thematic similarity, difference, and membership in three
  online mental health communities from reddit: a text mining and visualization
  approach.
\newblock {\em Computers in Human Behavior}, 78:98--112, 2018.

\bibitem{paul2016social}
M.~J. Paul, A.~Sarker, J.~S. Brownstein, A.~Nikfarjam, M.~Scotch, K.~L. Smith,
  and G.~Gonzalez.
\newblock Social media mining for public health monitoring and surveillance.
\newblock In {\em Proceedings of the Pacific symposium on Biocomputing}, pp.
  468--479. World Scientific, 2016.

\bibitem{perrotta2019spatio}
D.~Perrotta, D.~Delle~Vedove, C.~Obi, R.~Pebody, R.~Schifanella, and
  D.~Paolotti.
\newblock Spatio-temporal analysis of flu-related drugs uptake in an online
  cohort in england.
\newblock In {\em {Proceedings of the International Conference on Digital
  Public Health}}, pp. 111--118, 2019.

\bibitem{pick2019emotional}
S.~Pick, L.~H. Goldstein, D.~L. Perez, and T.~R. Nicholson.
\newblock Emotional processing in functional neurological disorder: a review,
  biopsychosocial model and research agenda.
\newblock {\em Journal of Neurology, Neurosurgery \& Psychiatry},
  90(6):704--711, 2019.

\bibitem{plutchik2001nature}
R.~Plutchik.
\newblock The nature of emotions: Human emotions have deep evolutionary roots,
  a fact that may explain their complexity and provide tools for clinical
  practice.
\newblock {\em American scientist}, 89(4):344--350, 2001.

\bibitem{ricci2018obesity}
G.~Ricci, D.~Tomassoni, I.~Pirillo, A.~Sirignano, M.~Sciotti, S.~Zaami, and
  I.~Grappasonni.
\newblock Obesity in the european region: social aspects, epidemiology and
  preventive strategies.
\newblock {\em Eur Rev Med Pharmacol Sci}, 22(20):6930--6939, 2018.

\bibitem{geo-big-data}
A.~C. Robinson, U.~Demšar, A.~B. Moore, A.~Buckley, B.~Jiang, K.~Field, M.-J.
  Kraak, S.~P. Camboim, and C.~R. Sluter.
\newblock Geospatial big data and cartography: research challenges and
  opportunities for making maps that matter.
\newblock {\em International Journal of Cartography}, 3(sup1):32--60, 2017.

\bibitem{robinson2019measuring}
P.~Robinson, D.~Turk, S.~Jilka, and M.~Cella.
\newblock Measuring attitudes towards mental health using social media:
  investigating stigma and trivialisation.
\newblock {\em Social psychiatry and psychiatric epidemiology}, 54(1):51--58,
  2019.

\bibitem{dataurban_4_2019}
J.~Schwabish.
\newblock {4 {Observations} on {Animating} {Your} {Data} {Visualizations}}.
\newblock
  \url{https://medium.com/@urban_institute/4-observations-on-animating-your-data-visualizations},
  2019.
\newblock [Online; accessed Aug. 2019].

\bibitem{segel_narrative_2010}
E.~{Segel} and J.~{Heer}.
\newblock Narrative visualization: Telling stories with data.
\newblock {\em IEEE Transactions on Visualization and Computer Graphics},
  16(6):1139--1148, 2010.

\bibitem{suler2004online}
J.~Suler.
\newblock The online disinhibition effect.
\newblock {\em Cyberpsychology \& Behavior}, 7(3):321--326, 2004.

\bibitem{suls2004evolution}
J.~Suls and A.~Rothman.
\newblock Evolution of the biopsychosocial model: prospects and challenges for
  health psychology.
\newblock {\em Health psychology}, 23(2):119, 2004.

\bibitem{tutubalina2017combination}
E.~Tutubalina and S.~Nikolenko.
\newblock Combination of deep recurrent neural networks and conditional random
  fields for extracting adverse drug reactions from user reviews.
\newblock {\em Journal of Healthcare Engineering}, 2017, 2017.

\bibitem{van2016biopsychosocial}
L.~Van~Oudenhove, R.~L. Levy, M.~D. Crowell, D.~A. Drossman, A.~D. Halpert,
  L.~Keefer, J.~M. Lackner, T.~B. Murphy, and B.~D. Naliboff.
\newblock Biopsychosocial aspects of functional gastrointestinal disorders: how
  central and environmental processes contribute to the development and
  expression of functional gastrointestinal disorders.
\newblock {\em Gastroenterology}, 150(6):1355--1367, 2016.

\bibitem{scepanovic2020medred}
S.~\v{S}\'{c}epanovi\'{c}, M.-L. Enrique, D.~Quercia, and K.~Baykaner.
\newblock Extracting medical entities from social media.
\newblock In {\em {ACM} Conference on Health, Inference, and Learning}, p.~12.
  Association for Computing Machinery, 2020.

\bibitem{scepanovic_martin-lopez_quercia_2020}
S.~\v{S}\'{c}epanovi\'{c}, E.~Martín-López, and D.~Quercia.
\newblock Medred dataset.
\newblock \url{https://figshare.com/articles/dataset/MedRed/12039609/1}, Mar.
  2020.

\bibitem{wang2006visualization}
W.~Wang, H.~Wang, G.~Dai, and H.~Wang.
\newblock Visualization of large hierarchical data by circle packing.
\newblock In {\em Proceedings of the SIGCHI conference on Human Factors in
  computing systems}, pp. 517--520, 2006.

\bibitem{wilbanks2014review}
B.~A. Wilbanks and P.~A. Langford.
\newblock A review of dashboards for data analytics in nursing.
\newblock {\em CIN: Computers, Informatics, Nursing}, 32(11):545--549, 2014.

\end{thebibliography}

\end{document}